%% file: gain_article.tex
\documentclass[a4paper]{jpconf}

\usepackage[utf8]{inputenc}
\usepackage{amsmath}
\usepackage{amsfonts}
\usepackage{amssymb}
\usepackage{graphicx}
\usepackage{siunitx}
\usepackage{hyperref}
\usepackage{subcaption}

\usepackage{floatrow}
\newfloatcommand{capbtabbox}{table}[][\FBwidth]

\usepackage{textgreek}
\newcommand{\keff}{k_\mathrm{eff}}
\newcommand{\first}{1\textsuperscript{st} }
\newcommand{\second}{2\textsuperscript{nd} }
\newcommand{\abs}[1]{\ensuremath{\left\vert#1\right\vert}}

\begin{document}

\title{Mobile quantum gravity sensor with unprecedented stability}

\author{C. Freier,$^{1\ast}$ M. Hauth,$^{1}$ V. Schkolnik,$^{1}$ B. Leykauf,$^{1}$
		M. Schilling$^2$,H. Wziontek$^3$, H.-G. Scherneck$^4$
		J. M\"{u}ller$^{2}$ and A. Peters$^{1}$}
	
\address{$^1$ Humboldt Universit\"{a}t zu Berlin, Institut f\"{u}r Physik, Newtonstr. 15, 
		Berlin, Germany}
\address{$^2$ Inst. f\"{u}r Erdmessung, Leibniz Universit\"{a}t Hannover, 
		Schneiderberg 50, Hannover, Germany}
\address{$^3$ Federal Agency for Cartography and Geodesy, Leipzig, Germany}
\address{$^4$ Onsala Space Observatory, Chalmers University of Technology, Gothenburg, Sweden}
		
\ead{christian.freier@physik.hu-berlin.de, achim.peters@physik.hu-berlin.de}

\begin{abstract}
Changes of surface gravity on Earth are of great interest in geodesy, earth sciences and natural resource exploration. They are indicative of Earth system’s mass redistributions and vertical surface motion, and are usually measured with falling corner-cube- and superconducting gravimeters (FCCG and SCG).
Here we report on absolute gravity measurements with a mobile quantum gravimeter based on atom interferometry. The measurements were conducted in Germany and Sweden over periods of several days with simultaneous SCG and FCCG comparisons. They show the best-reported performance of mobile atomic gravimeters to date with an accuracy of \SI{39}{nm/s^2} and long-term stability of \SI{0.5}{nm/s^2}, short-term noise of \SI{96}{nm/s^2/\sqrt{Hz}}.
These measurements highlight the unique properties of atomic sensors. The achieved level of performance in a transportable instrument enables new applications in geodesy and related fields, such as continuous absolute gravity monitoring with a single instrument under rough environmental conditions.
\end{abstract}

Gravimetry today is based on both absolute and relative measurements\cite{crossley_measurement_2013}. Falling corner-cube absolute gravimeters (FCCG) track the ballistic trajectory of a free-falling test mass using a light interferometer to obtain the local gravitational acceleration. Relative gravimeters are based on a spring-mass system or a superconducting sphere levitating in a magnetic field to measure differential signals with respect to some reference value\cite{crossley_measurement_2013}. 
While all these instruments have individual merits in advancing the field of gravimetry, they also have characteristic shortcomings.
For instance, the best FCCGs\cite{niebauer_new_1995} have a small total uncertainty of \SI{20}{nm/s^2} but require sites with a low level of micro-seismic vibrations to reach their optimal performance \cite{vancamp_uncertainty_2005}. The moving internal test-mass furthermore excites floor vibrations and necessitates massive concrete foundations in order to reduce self-induced vibrations \cite{charles_vertical_1995}.
FCCG were also not built to perform continuous gravity monitoring over extended periods of time due to wear of their moving mechanical parts.
Long-term gravity registrations are therefore done with relative gravimeters. Because of their working principle these instruments depend on repeated calibration and drift-rate verification through an absolute gravity reference which is, in the case of SCG, provided by FCCG.
Relative and absolute gravity measurements are thus complementary techniques and must usually be used in combination.
As shown below, atomic gravimeters overcome many limitations of existing instruments through their ability to generate continuous, absolute gravity measurements over long periods of time with one mobile, stand-alone instrument.

Atom interferometers (AI) using light pulses as beam-splitters and mirrors for atomic matter waves were first realized in 1991\cite{riehle_optical_1991,kasevich_atomic_1991} and have since become a versatile technology not only for inertial sensing applications\cite{peters_highprecision_2001,gillot_stability_2014,tackmann_selfalignment_2012} but also for precision measurements of the Newtonian gravitational constant\cite{rosi_precision_2014}, the fine-structure constant\cite{wicht_preliminary_2002}, as well as for proposed tests of the equivalence principle\cite{dimopoulos_testing_2007} or future gravitational wave detectors\cite{dimopoulos_gravitational_2009}.

\section{The Gravimetric Atom Interferometer}
\begin{figure}[h]
	\centering
	\resizebox{0.8\textwidth}{!}{
		\includegraphics{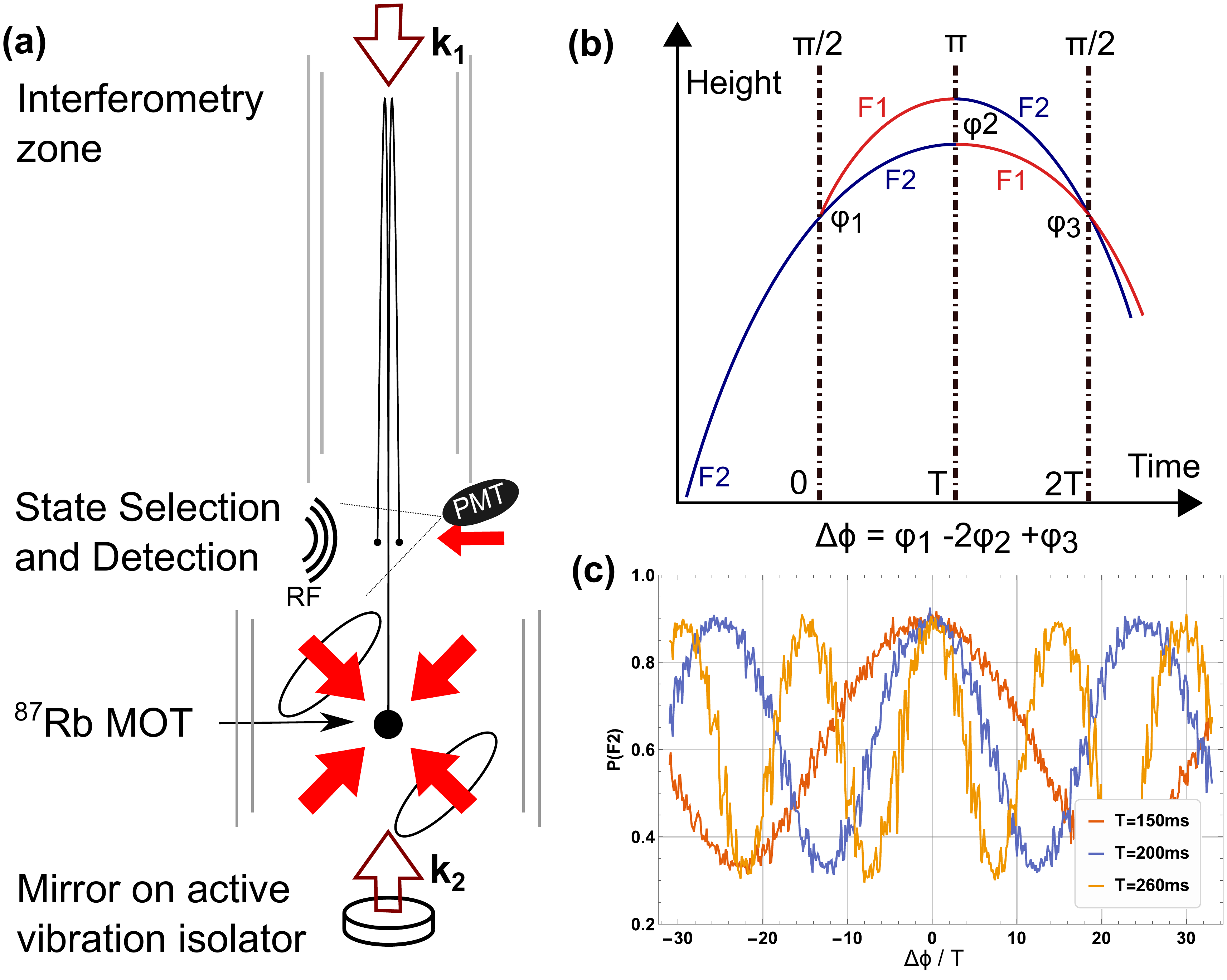}
	}
	\caption{\textbf{GAIN working principle}. \textbf{a}, atomic fountain set-up with magneto-optical trap (MOT) and Raman laser beams $\mathbf{k}_1$,$\mathbf{k}_2$ for atom interferometry (AI). \textbf{b}, Mach-Zehnder AI sequence under the influence of gravity. \textbf{c}, Interferometer fringes encoded in the internal state population (F2) for several values of $T$.}
	\label{fig:physicspackage} 
\end{figure}

The gravimetric atom interferometer (GAIN) was developed and built at Humboldt Universit\"{a}t zu Berlin to perform mobile high-precision gravity measurements. It is based on interfering ensembles of $^{87}$Rb atoms in an atomic fountain configuration and stimulated Raman transitions.
Fig. \ref{fig:physicspackage} shows the working principle of the instrument, a complete description has been published elsewhere\cite{hauth_atom_2014,hauth_first_2013}.
During free fall the atomic matter-waves are split vertically in space, reflected and recombined using a Mach-Zehnder $\frac{\pi}{2}$-$\pi$-$\frac{\pi}{2}$ pulse sequence, compare Fig. \ref{fig:physicspackage}(b).
The output phase of the interferometer depends on the local light phase at the atomic positions during the pulses. This corresponds\cite{hauth_atom_2014} to a threefold sampling of the ballistic trajectory and thus a $g$ measurement, when neglecting the spatial splitting of the wave-packets during the interferometry sequence.
The measurement sequence starts with a Magneto-Optical Trap (MOT) loading approximately $10^9$ laser-cooled $^{87}$Rb atoms within \SI{0.6}{s}. 
The cloud is then launched upwards using far-detuned moving molasses achieving a temperature of \SI{2}{\micro K} and a vertical velocity of \SI{4}{m/s}.
After a previously described process\cite{hauth_atom_2014} which selects only atoms in the magnetically insensitive sub-state within a narrow vertical velocity class, the cloud reaches the elongated magnetically shielded interferometry region. 
The $\frac{\pi}{2}$-$\pi$-$\frac{\pi}{2}$ Raman pulse sequence is then conducted with a time $T=\SI{0.26}{s}$ between pulses. Afterwards, the upper state population $P_{2}$ is detected using a fluorescence detection system which yields an interference signal $P_{2}=\bar{P}+ C/2 \cos{\Delta\Phi}$ with contrast $C$ and offset $\bar{P}$. The next measurement is then started with a cycle time of \SI{1.5}{s}.
The interferometer phase  $\Delta \Phi$ is connected to the gravity value $g$ by \(\Delta \Phi = \left (\keff g - \alpha \right )T^2 +\Phi_L\), where $\mathrm{k_{eff}}=\abs{\mathbf{k}_1 -\mathbf{k}_2} \approxeq 2 \mathrm{k}$ is the effective wave-number of the Raman transitions, $\alpha$ is a constant frequency chirp rate of the Raman laser which compensates for the time-varying Doppler-shift of the atoms during free fall, and $\Phi_L$ is a configurable Raman laser phase which can be used to scan the interferometer fringe.
In order to deduce $\Delta \Phi$ at a new measurement site, interference fringes are first scanned by altering the RF chirp rate $\alpha$ for several values of $T$. Their positions overlap at the desired chirp rate value where $\alpha=\keff g$ which removes the phase ambiguity due to the fringe pattern, compare above equation and Fig. 1(c).

The inertial reference is provided by a retro-reflection mirror shown in the bottom of Fig. \ref{fig:physicspackage}(a) which therefore has to be decoupled from environmental vibrations.
This is realised with an active vibration isolation system\cite{hauth_atom_2014}. Due to its limited isolation at micro-seismic frequencies of \SIrange[range-phrase=--,range-units=single]{0.1}{0.3}{Hz}, residual vibrations are contained in its feedback loop error signal. Feeding this signal into a previously demonstrated post-correction algorithm\cite{legouet_limits_2008} thus further reduces interferometer phase noise by a factor 2 to 4, depending on the micro-seismic background.
Special care was taken during the development of GAIN to ensure that the set-up is small and robust enough for transport to other locations with minimal effort. It thus consists of 3 modular units\cite{hauth_atom_2014} with a respective size of roughly \SI[product-units = brackets-power]{1 x 1 x 2}{\metre}.
\section{Mobile gravity comparison campaigns}
After completion of the set-up and initial gravity comparisons in Berlin\cite{hauth_atom_2014}, two measurement campaigns were carried out at remote locations. The \first took place in November 2013 at the Geodetic Observatory in Wettzell, the \second in February 2015 at Onsala Space Observatory near Gothenburg, Sweden.
Both locations were chosen because the local gravity values are known very accurately due to several measurement campaigns conducted with FCCG during the last years\cite{timmen_observed_2015}.
They furthermore operate SCGs for high-precision monitoring of temporal gravity changes which proved invaluable for characterizing the performance of the GAIN gravimeter. 
While conducting the \second campaign in Onsala, the state-of-the-art FCCG FG5X-220 performed a simultaneous measurement for a direct comparison of the instrumental properties.
During both campaigns a small truck was used to transport GAIN to the destination sites. Transport and set-up took approximately 3 days until the first gravity data were produced, a delay which could be shortened in the future to less than a day through optimization of the work-flow and small alterations of GAIN.
\begin{figure}[h]
	\begin{subfigure}[b]{0.495\textwidth}
		\resizebox{\textwidth}{!}{
			\includegraphics{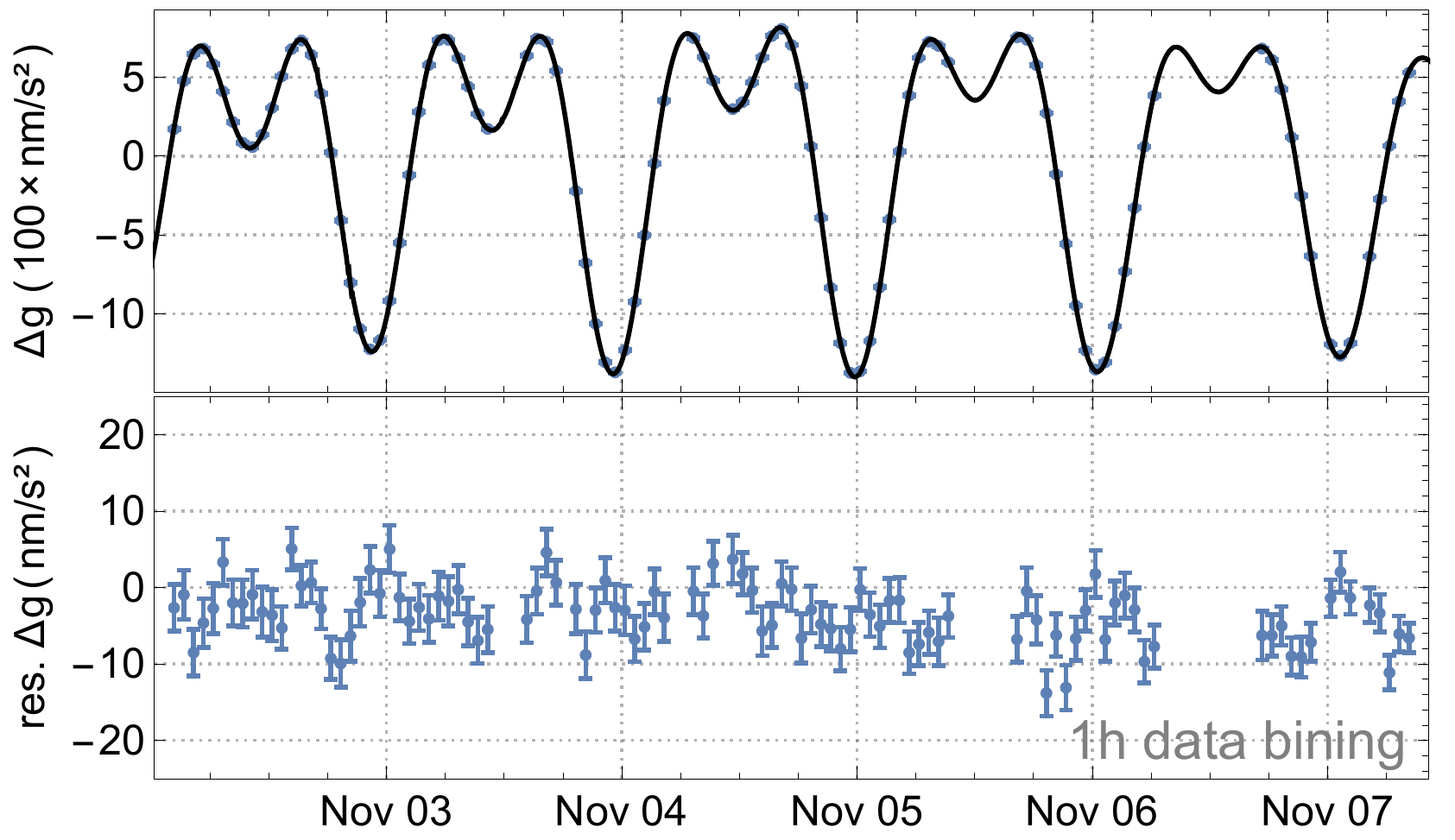}
		}
		\caption{Wettzell, Germany (2013)}
	\end{subfigure}	
	\begin{subfigure}[b]{0.495\textwidth}
		\resizebox{\textwidth}{!}{
			\includegraphics{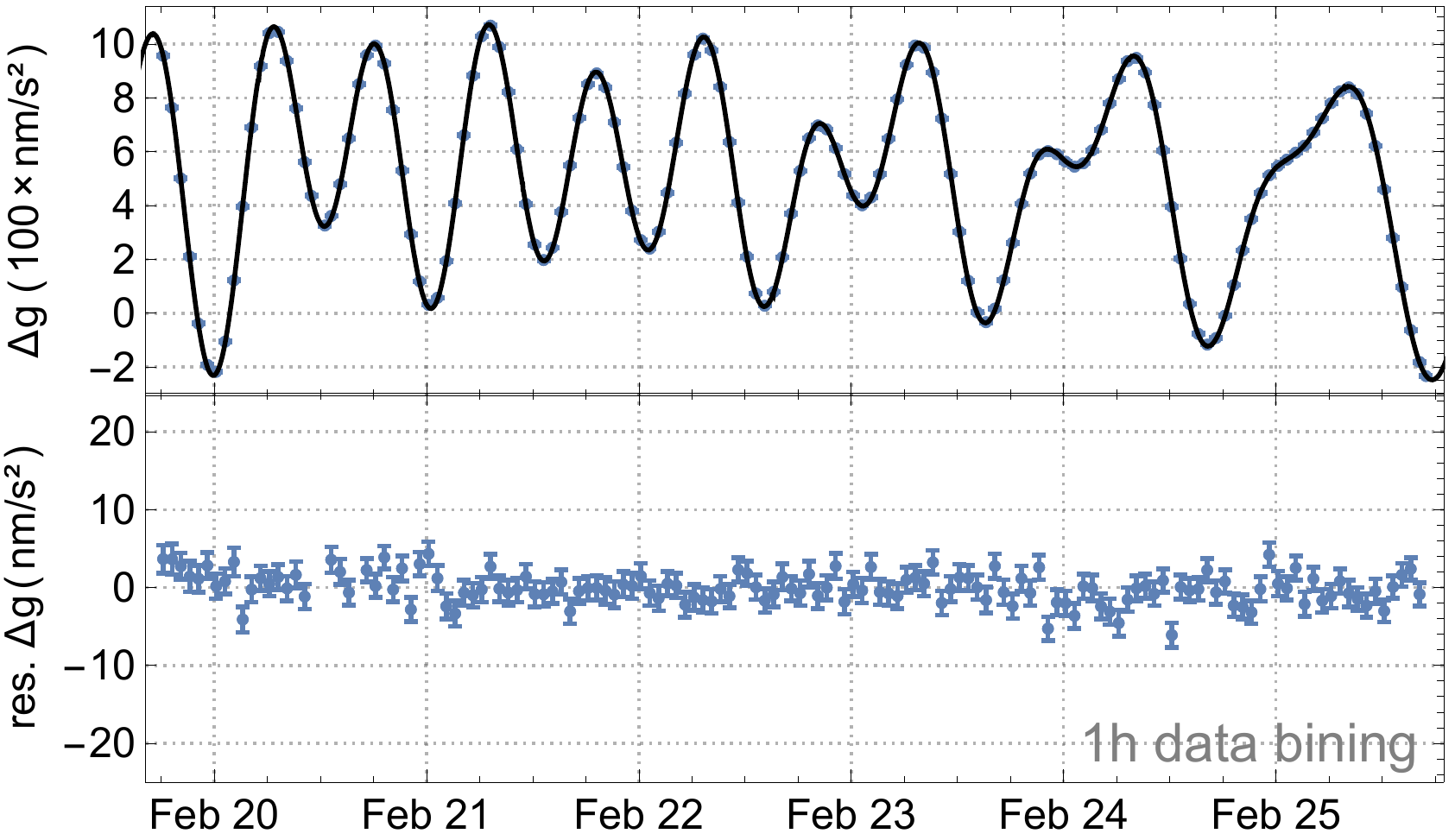}
		}
		\caption{Onsala, Sweden (2015)}
	\end{subfigure}
	\caption{\textbf{GAIN gravity data.} \textbf{a}, Top: tidal signal measured by GAIN (blue) versus SCG (black) during the \first campaign. Bottom: difference between GAIN and the SCG signals. \textbf{b}, equivalent plot of \second campaign. The decreased noise level and improved stability during the \second campaign is due to enhancements of GAIN between both campaigns.}
	\label{fig:gravitydata}
\end{figure}

\section{Gravimeter Performance}
\begin{figure}[h]
	\centering
	\begin{subfigure}[b]{0.495\textwidth}
		\resizebox{\textwidth}{!}{
			\includegraphics{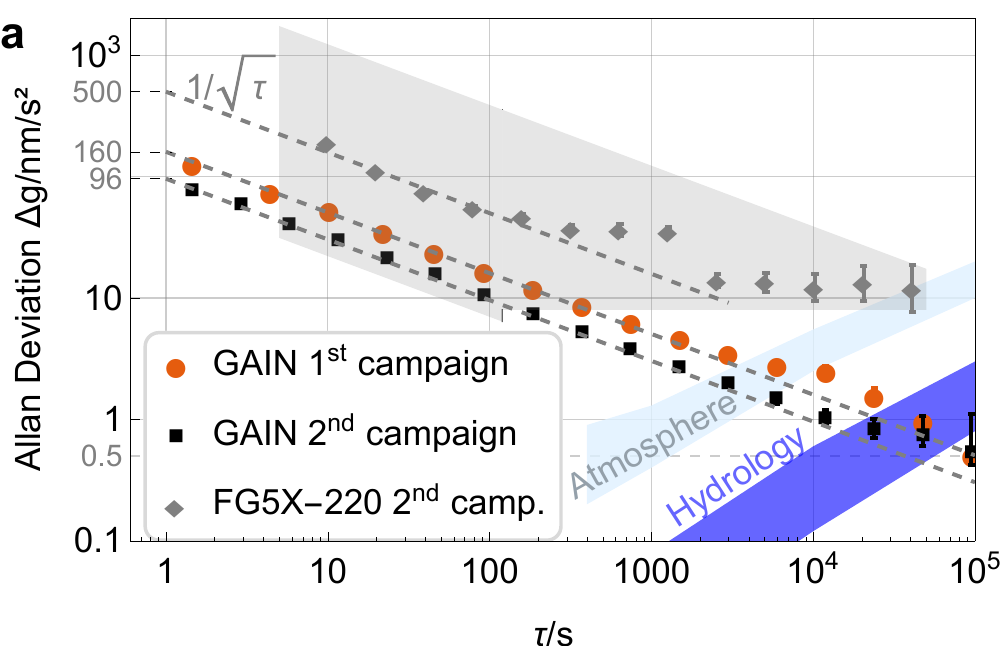}
		}
	\end{subfigure}
	\begin{subfigure}[b]{0.495\textwidth}
		\resizebox{\textwidth}{!}{
			\includegraphics{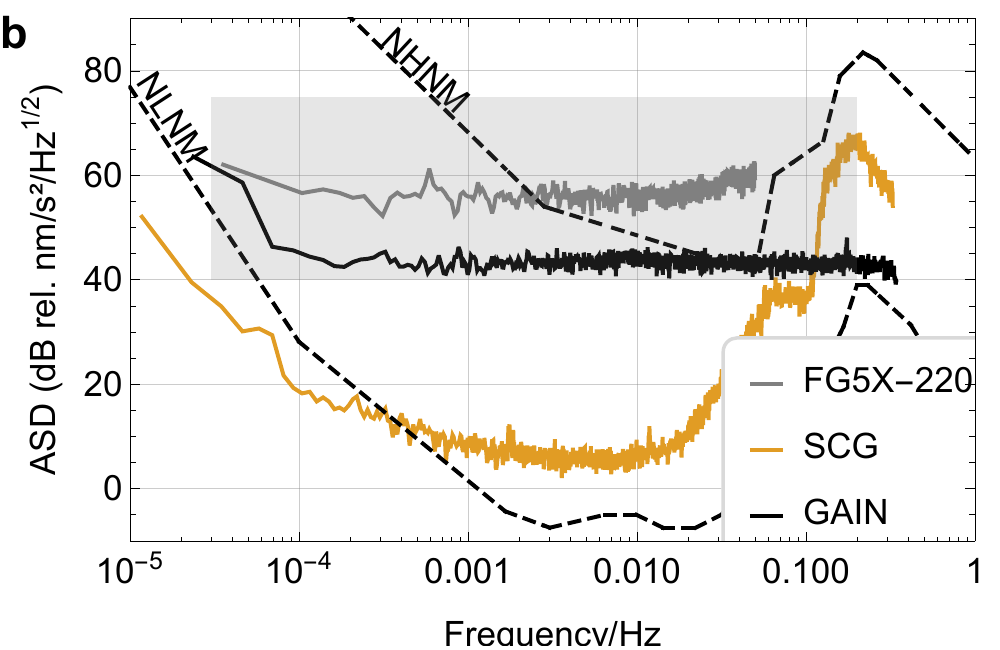}
		}
	\end{subfigure}
	\caption{\textbf{Gravimeter comparison.} \textbf{a}, Allan std. deviation of GAIN and FG5X-220 gravity data reduced by SCG signals. The continuous operation of GAIN allows shorter integration time and improved long-term stability compared to the FG5X. 
	The blue-shaded areas highlight common gravity signals due to air-pressure and hydrological effects.
	\textbf{b}, \textbf{Amplitude Spectral Density} of GAIN during the \second campaign, showing a noise level better than the FG5X but worse than SCG.
	The FG5X noise level was increased due to elevated micro-seismic vibrations as depicted by high- and low-noise model\cite{peterson_observations_1993}. The gray areas thus denotes the expected noise under quiet and noisy conditions\cite{vancamp_uncertainty_2005}. 
	}
	\label{fig:allandev}
\end{figure}

A five day segment of the gravity data taken during both campaigns is shown in the upper half of Fig. \ref{fig:gravitydata}, showing temporal variations due to earth tides, atmospheric and other environmental effects \cite{crossley_measurement_2013}.
In order to characterize the gravimeter's stability, these variations need to be subtracted from the gravity signal. Below a magnitude of \SI{10}{nm/s^2} it is not sufficient to use only synthetic models for this purpose because remaining signals caused by incomplete atmospheric correction, tidal modelling and local water storage fluctuations can reach up to a few \SI{10}{nm/s^2}. 
We thus use the SCG as a reference to assess instrumental effects in GAIN. As shown in the bottom half of Fig. \ref{fig:gravitydata}, both instrument are in excellent agreement.
The residuals obtained by subtracting both datasets do not contain visible signals within the error bars of the measurement. 
The detailed analysis in Fig. \ref{fig:allandev} shows that they resemble white noise on time-scales below \SI{e4}{s}. 
Beyond this the relative stability improves further and reaches \SI{0.5}{nm/s^2}, or \SI{5e-11}{g}, for a time-scale of \SI{e5}{s} or approximately one day.
This is the best reported stability for an absolute gravimeter to our knowledge and shows the excellent control of any time-variable systematic effects in GAIN as listed in table \ref{tab:systematics}.
Comparing SCG and GAIN data on the sub-\si{nm/s^2} level requires SCG scale factor calibration significantly better then \num{e-3} which is beyond standard FCCG results\cite{francis_evaluation_2002}.
The scale factor was thus re-obtained for both campaigns using 5 days of GAIN gravity data, yielding values consistent with previous calibrations but improved uncertainties of \num{4e-4} and \num{2.6e-4}, respectively. 
Even under optimized schedules\cite{vancamp_optimized_2015}, low micro-seismic noise and maximal tidal amplitude, FG5 instruments would need at least three weeks to achieve a similar precision which demonstrates the value of GAIN's improved resolution.

The absolute gravity values were compared to the reference value during the \first campaign and to the simultaneously measured FG5X-220 result during the \second campaign. 
Both comparisons show excellent agreement within their specified $1\sigma$ uncertainties as shown in table \ref{tab:comparison}.
The GAIN values are slightly higher in both cases with differences of \SI[separate-uncertainty = true,tight-spacing = true]{62\pm64}{nm/s^2} and \SI[separate-uncertainty = true,tight-spacing = true]{32\pm39}{nm/s^2}.
The achieved level of accuracy is slightly better than another value \cite{francis_ccmgk2_2015} obtained using atom interferometry and comparable to the level provided by the FG5(X), where individual instruments differ by \SIrange[range-phrase=--,range-units=single]{20}{30}{nm/s^2} \cite{crossley_measurement_2013,francis_ccmgk2_2015}.
The biggest uncertainty is caused by Raman laser wave-front distortions as shown in table \ref{tab:systematics}.
They have been reduced by wave-front measurements and a correcting model \cite{schkolnik_effect_2015} which was applied to all optical elements in the Raman beam path except the lower vacuum window. 
In the future this effect could be decreased further by reducing the cloud expansion through advanced cooling techniques or a more complete correction.
\begin{figure}
\begin{floatrow}
	\capbtabbox{
		\begingroup
		\setlength{\tabcolsep}{1pt} 
		\begin{tabular}{ l c c }
			\br
			\textbf{Systematic effect} & \textbf{Bias}  & \textbf{Error} \\
			&\multicolumn{2}{c}{\textbf{(nm/s$^2$)}}\\
			\mr
			Raman Wavefronts & -28 & $\pm$22\\ 
			Coriolis Effect& 0 & $\pm$15 \\
			Magnetic Field Effects & 0 & $\pm$10 \\	
			RF Group delay & 0 & $\pm$10 \\ 
			Self Gravitation & 19 & $\pm$5 \\ 
			Reference Laser Freq. & -12(-10) & $\pm$5 \\ 
			Synchronous Vibrations& 0(90) & $\pm$5(50) \\ 
			AC Stark Shift (1PLS) & 0 & $\pm$5 \\ 
			Rb background vapour & 5 & $\pm$3 \\ 
			AC Stark Shift (2PLS) & 0 & $\pm$2 \\ 
			vertical alignment & 0(-1) & $\pm$1 \\ 
			\hline
			\textbf{Total} & \textbf{-16(77) } & \textbf{$\pm$32(61)} \\			
			\br 
		\end{tabular}	
		\endgroup
	}
	{
		\caption{\textbf{GAIN systematic error budget} for the \second campaign. The total bias was subtracted from GAIN raw value to obtain the corrected result. Values for the \first campaign are denoted in brackets. Synchronous vibrations were eliminated between campaigns by means of a magnetic shield around the MOT region.}
		\label{tab:systematics}
	}
	\capbtabbox{
	\begin{tabular}{ l r }
		\br
		\textbf{First campaign} & \textbf{Gravity (nm/s$^2$)}\\
		GAIN raw value & 9,808,369,362 $\pm$ 01\\ 
		Systematic bias & 77 $\pm$ 61 \\
		Height offset & 400 $\pm$ 10 \\
		GAIN corrected & 9,808,369,685 $\pm$ 62\\
		Reference value & 9,808,369,623 $\pm$ 18\\	
		\textbf{GAIN-Ref.} & \textbf{62 $\pm$ 64}\\ 
		\br
		\multicolumn{2}{l}{\textbf{Second Campaign}}\\
		GAIN raw value & 9,817,158,312 $\pm$ 01 \\ 
		Systematic bias & -16 $\pm$ 32\\
		Height offset & 727 $\pm$ 10\\
		GAIN corrected & 9,817,159,055 $\pm$ 34\\
		FG5X-220 value & 9,817,159,023 $\pm$ 20\\
		\textbf{GAIN-FG5X} & \textbf{32 $\pm$ 39}\\ 		
		\br 
	\end{tabular}	
	}{
	\caption{\textbf{Absolute gravity comparison} of the atomic gravimeter and known gravity reference values during both campaigns, yielding excellent agreement. The vertical gravity gradient was measured independently beforehand and used to account for effective measurement height differences between values.}
	\label{tab:comparison}
	}
\end{floatrow}
\end{figure}
\section{Discussion}
Due to stormy winter weather and the proximity to the Kattegat coastline, micro-seismic vibrations were strongly elevated during parts of the \second campaign. This did not affect the GAIN sensor as shown in Fig. \ref{fig:allandev}(b), but increased the FG5X noise level roughly five-fold as observed before\cite{vancamp_uncertainty_2005,hauth_atom_2014,gillot_stability_2014}.
GAIN's improved tolerance to micro-seismic vibrations is due to its \SI{0.7}{Hz} repetition rate, as opposed to \SI{0.1}{Hz} of the FG5X, and more effective isolation system. This shows a general advantage of atomic gravimeters compared to FCCGs which rely on low levels of micro-seismic excitation. 
Atomic sensors also do not excite vibrations of the floor/pillar as the test-masses involved in the measurement are negligible, which altogether enables low-noise operation not only on dedicated concrete pillars but also on regular flooring.
This extends their range to a vastly increased number of measurement sites which, in combination with their continuous, drift-free, absolute measurement, will improve results in applications such as monitoring of tidal, hydrological or geophysical gravity signals with amplitudes in the \SIrange[range-units = single]{1}{10}{nm/s^2} range on long time-scales where spring-based instruments are limited by non-linear drifts.
Albeit GAIN can not yet reach the ultimate low short-term noise level provided by SCG, the values demonstrated here are comparable to those of state-of-the-art spring-based relative gravimeters \cite{riccardi_comparison_2011} without their long-term drifts. 
The noise level of mobile atomic gravimeters may in the future be improved through further optimization of vibration isolators\cite{hu_demonstration_2013}, hybrid sensor strategies\cite{lautier_hybridizing_2014} or advanced techniques such as interleaved fountain operation\cite{meunier_stability_2014}.
The comparison between two types of absolute gravimeters based on fundamentally different measurement principles, namely FCCG and atomic gravimeters, is beneficial in its own right and will boost the search for hidden systematic effects and increase the accuracy of today's gravity networks.
We further anticipate that the robust and adaptable architecture of cold atom interferometry will enable the development of self-contained, compact and rugged instruments for applications which so far were not accessible to absolute gravimetry.
For instance, airborne and shipborne gravity surveys will benefit from the characteristics of atomic gravimeters and gradiometers due to their intrinsic immunity to hysteresis effects which deteriorate the performance of current spring-based instruments during turbulence\cite{forsberg_airborne_2010}.
Drift-free, continuous operation at places such as underground systems, boreholes or even the sea floor will exceed the operating range of existing spring-based gravimeters and enable new insights in earth sciences and geodesy.

\ack{This material is based on work funded by the European Commission (FINAQS, Contr. No. 012986-2 NEST), by ESA (SAI, Contr. No. 20578/07/NL/VJ) and by ESF/DFG (EuroQUASAR-IQS, DFG grant PE 904/2-1 and PE 904/4-1).}

\section*{References}
\input{gain_article.bbl}

\end{document}

%% file: gain_article.bbl
\providecommand{\newblock}{}